\documentclass[graybox]{svmult}

\usepackage{type1cm}        
\usepackage{makeidx}         
\usepackage{graphicx}        
\usepackage{multicol}        
\usepackage[bottom]{footmisc}

\usepackage{newtxtext}       %
\usepackage[varvw]{newtxmath}       
\usepackage{aas_macros-3}

\makeindex             

\begin{document}

\title*{General History of X-Ray Polarimetry in Astrophysics}
\author{Enrico Costa}
\institute{Enrico Costa \at Istituto di Astrofisica e Planetologia Spaziale - INAF, Via del Fosso del Cavaliere, 00133 Roma. Italy, \email{enrico.costa@inaf.it}}
%
%
\maketitle

\abstract*{Soon after the discovery of the first extrasolar X-Ray sources it was suggested that polarimetry could play a major role as a diagnostic tool. Attempts to measure polarization of X-Ray sources was performed by the team of Columbia University lead by Robert Novick. The technique of Bragg diffraction at $45^{0}$ was successful to detect the polarization of the Crab with rockets and with OSO-8 satellite. In the following evolution of X-Ray Astronomy, Polarimetry was too mismatched with the improved sensitivity of imaging and spectroscopy, based on the use of optics. As a consequence no polarimeter was flown any more.
At the beginning of the century a new class of instruments based on the photoelectric effect were developed. In the focus of an X-Ray telescope they can perform angular, energy and time resolved polarimetry and benefit of the large increase of sensitivity due to the optics.
The Imaging X-Ray Polarimetry Explorer, exploiting this technique, was launched at the end of 2021.}

\abstract{Soon after the discovery of the first extrasolar X-Ray sources it was suggested that polarimetry could play a major role as a diagnostic tool. Attempts to measure polarization of X.Ray sources was performed by the team of Columbia University lead by Robert Novick. The technique of Bragg diffraction at $45^{0}$ was successful to detect the polarization of the Crab with rockets and with OSO-8 satellite. In the following evolution of X-Ray Astronomy, Polarimetry was too mismatched with the improved sensitivity of imaging and spectroscopy, based on the use of optics. As a consequence no polarimeter was flown any more.
At the beginning of the century a new class of instruments based on the photoelectric effect were developed. In the focus of an X-Ray telescope they can perform angular, energy and time resolved polarimetry and benefit of the large increase of sensitivity due to the optics.
The Imaging X-Ray Polarimetry Explorer, exploiting this technique, was launched at the end of 2021.}

\section{The very early stage}
\label{sec:1}

Only 9 years after the discovery of X-Rays by Roentgen in 1904 Charles Glover Barkla, a student of Stokes, made experiments of scattering of these newly discovered particles and found that they follow the same rules of polarization of optical light and this demonstrated that these mysterious X-rays were electromagnetic radiation. Physics and Polarimetry of X-Rays are born almost together. 
When in 1962 Giacconi and Rossi found the first evidence for extrasolar X-ray sources theoreticians predicted that polarimetry would have in this wavelength domain a major role than in other domains. The Soviet Physicist Vitali Ginzburg was very active in this promotion also toward the western community.
The easiest implementation of this concept would be an instrument based on the angular distribution of photons scattered from a target. The difficulty is that the scattering prevails over the absorption at energies of few keV, where the majority of photons is, only with Hydrogen and Helium, that have a negligible stopping power for X-Ray photons. The best dense scattering material is Lithium (that in any case must be encased with beryllium) for which the scattering prevails only above 10 keV.
In 1969 Herbert Schnopper suggested the use of Bragg crystals at incidence angles of $45^{0}$ as a good analyzer of linear polarization\cite{Schnopper1969}. The technique would be very robust and simple. The limit is the narrow band of the diffracted radiation that would result in a low effective area. Moreover the detector would be of the same surface of the entrance window, and the poor resolution of the detectors would not allow to separate effectively by the pulse height the diffracted photons, so that a large background could be expected.

The team of Columbia University lead by Robert Novick carried on a systematic study to arrive to perform an astronomical experiment. His co-workers Roger Angel and Martin Weisskopf found that the efficiency could be improved by using mosaic crystals of pyrolytic graphite (i.e. a crystal with the spread of orientation of micro-domains artificially enhanced). Moreover they found that a moderate bending of the crystal would allow a concentration of diffracted photons on a small detector, with a much lower background rate, while substantially preserving the modulated response to polarization\cite{Angel1970},\cite{Weisskopf1972}. 

The Team of Columbia launched the first rocket searching for polarization of ScoX-1 on 1968\cite{Angel1969}. The rocket included scattering targets of lithium encased in beryllium surrounded by proportional counters. The polarization angle was derived from the angle between the scatterer and the detector and the spinning phase of the rocket. The observation of ScoX1 (performed only 6 years after the discovery of the source!) was unsuccessful in terms of polarimetry. But the same rocket payload, was launched again to point the Crab Nebula. The result was inconclusive but still compatible with the high polarization found in optical and radio bands. When the technique of Bragg diffractors wih mosaic graphite was mature an improved version of the rocket was built. Out of the atmosphere four panels mounting graphite crystals were deployed on the side at an angle of $45^{0}$  to the pointing/spinning axis of the rocket so that the photons parallel to the axis were diffracted toward proportional counters hosted in the rocket below the scattering stage of the payload that was similar to that of the first rockets.  Also this result was inconclusive. But by overimposing the data of all flights Novick found at last a statistically significant evidence of polarization\cite{Novick1972}.

This was a paramount result in terms of physics of X-ray sources also confirming the diagnostic relevance of this subtopics, notwithstanding the experimental difficulties. Another achievement for the future was that, in the few keV range, the Bragg approach, notwithstanding the very low efficiency, is better than Thomson in terms of both sensitivity and reliability 

While performing this first set of experiments the Novick Team also fixed the statistic frame that would be used for all the planned and the (few) performed experiments. Both scattering and Bragg polarimeters would produce a histogram of phase of emerging photons. The histogram can be fitted with a constant term plus a cos$^{2}$ term. The ratio between the two terms carries the information about polarization degree while the phase of the second term carries the information n the polarization angle.
\begin{equation}
M=\dfrac{S_{max}-S_{min}}{S_{max}+S_{min}}
\end{equation}
The M is named modulation. The modulation for a beam 100$ \% $ polarized, conventionally named modulation factor ($\mu$) is a feature of the detector and usually depends on the energy. If counts follow Poisson Statistics, as usually do in X-ray detectors, the modulation, which is a positively defined quantity, follows a distribution of $ \chi^{2} $ for two degrees of freedom. This is the basis to define the significance of a possible detection. The Minimum Detectable Modulation can be computed starting from the statistics as the modulation that in only 1$\%$ of occurrences can be arrived or passed by statistical fluctuation. By dividing for the modulation factor we can derive the Minimum Detectable Polarization (MDP) that is the usual parameter to describe the sensitivity of an experiment of X-ray polarimetry when studying a source of a certain flux and spectrum. 
\begin{equation}
MDP=\dfrac{4.29}{\mu\varepsilon S}\times\sqrt{\dfrac{\varepsilon S+B}{T}}
\end{equation}

The fact that a 99$ \% $ confidence was chosen, instead than the most usual 3$ \sigma $ (or 5$ \sigma $), witnesses the awareness that polarimetry would be a tough challenge. 

The Columbia team was also active in optical polarimetry of sources of potential interest for X-ray astronomy, so that it was familiar with the Stokes parameters formalism. Actually the results of the first rockets were presented and discussed in this formal frame, although both Bragg and Thomson stages can only detect linear polarization. The V parameter is unmeasurable so that the use of the Stokes is, in the frame of X-ray polarimetry only, of moderate usefulness. To compare results in a sky map a display of the polarization degree and angle could be the most informative. But in order to insert the X-ray result in a general frame of measurements in other wavelength or compare with theories the Stokes parameter representation would be the most suited.   In the topical paper where polarization is detected the discussion is concentrated on the technique to combine data of different experiments and flights and the related systematics, so that, in this case, the Stokes Parameters were ignored.    

\section{Ariel-5 and OSO-8}
\label{sec:2}
The first two satellites with an X-ray polarimeter aboard, Ariel-5 and OSO-8, and the only ones for the following 45 years, were launched on 1974 and 1975 respectively, only 5 and 6 years after the first X-Ray survey mission UHURU. The ARIEL-5 spectro-polarimeter designed and built by the Team of Leicester University lead by Richard Griffiths\cite{Griffiths1976} was also based on Bragg Diffraction but was somehow different. The direction of the radiation impinging on the instrument and that of the diffracted photons impinging on the detectors were defined by mechanical collimators as well. But the diffracting crystals were of two kinds, one of mosaic graphite, like OSO-8, the second one of LiF. Crystals were flat and their inclination, with respect to the incoming radiation could be adjusted in order to tune the mounting angle and the related diffraction energy. The spinning satellite would be pointed not to the source but to a direction slightly off-set. By this trick the photons from the source would impinge on the crystal with an angle slightly different at each phase of spin. This would allow to perform high resolution spectroscopy of expected Si and S lines. The LiF was used to make spectra of Fe lines. At an angle of $45^{0}$  the diffractometer could be used as a polarimeter around an energy of 2.5 keV. 

In OSO-8 experiment, lead by Novick and Weisskopf, both functions of spectroscopy and polarimetry were present but devoted to two dedicated instruments, both designed and built from the Team of Columbia University. The spectrometer was based on a flat crystal of pyrolytic graphite with the unavoidably large detectors. The polarimeter was based on graphite mosaic crystals, bent to concentrate photons on proportional counters. The diffraction angles were confined in a range around $45^{0}$ from mechanical collimators. The whole was in continuous rotation around the pointing axis through the satellite spinning. 
The eventual yield of both spectrometers was very poor. Only in one case  a line was detected. Upper limits on narrow lines were fixed of doubtful significance given that the observed sources were likely thick. The real effect was to discourage the future instruments of high resolution spectroscopy.

The two polarimeter had different results. Both OSO-8 and Ariel-5 used proportional counters with pulse shape and pulse height discrimination of background. The areas of the two instruments were comparable but OSO-8 would be much more sensitive. Thanks to the bent geometry the detectors of OSO-8 had a surface around 20 times smaller than the crystal, with an almost proportional reduction of the background. In the Leicester experiment the main use for spectroscopy forced the flat geometry and as a consequence a much higher background. Moreover the mosaic angular spread of the graphite was smaller than that of OSO-8, in order to preserve the spectral resolution and this means a lower efficiency for polarimetry. Last but not least an important fraction of the surface was used for the LiF crystal, which had an efficiency much lower than that of graphite.

OSO-8 made a long pointing of the Crab. Confirmed the results of the rockets measurement but with a very high significance (19$ \sigma $)\cite{Weisskopf1976}. The measured polarization was significantly above the so called Chandrasekhar limit (15.7$\%$ vs 12$\%$), namely the maximm polarization deriving from scattering on an asymmetric geometry. This was the final evidence of a not thermal component of the X-ray emission, likely synchrotron. 

A strong limit to ScoX1 polarization was also found\cite{Long1979}. But other results were of much lower impact. These results were in general considered disappointing. While the data on Crab rose a high interest, data on other sources were still consistent with theories but demonstrated that performing polarimetry was difficult, complex and technically cumbersome. 

But the main trouble with polarimetry arrived from the great transformation that was occurring in X-Ray Astronomy  in those years. Soon after the first discovery of extrasolar sources, Giacconi outlined a long term planning that foresaw, after the first explorer mission (that would be UHURU), a first medium size mission based on an X-ray telescope with, in the focal plane a revolver capable to alternate different instruments in the focus. This mission would become the  Einstein Mission\cite{Giacconi1979}. The High Resolution Imager produced impressive images of a limited sample of extended sources. But in practice the Imaging Proportional Counter performed the large majority of the work for point-like sources. The telescope collected the photons impinging on an area of 400 $cm^{2}$ and concentrated on a focal spot of the order of one $mm^{2}$  with the consequence that the background count was equivalent to that of a very weak source.  The increase in sensitivity for the detection of a source was impressive.  

With an observation of few hours Einstein/IPC could detect a source at the limit of UHURU sensitivity. Einstein disclosed the world of extragalactic sources that had been only
touched with collimated missions. 

The mismatching in sensitivity between imaging and polarimetry increased enormously. Moreover the viable polarimeters implied a rotation of the instrument, easily achieved with spinning satellites. On the contrary the new imaging/spectra instruments in the focus of the optics did not need any more rotation and a three axis pointing was much more useful for the measurement itself, for the use of power resources and for the pointing flexibility. Therefore a polarimeter would be a rotating equipment harbored in the focal plane. In other terms the addition of a polarimeter would represent a serious increase of complexity and a heavy share of observing time to make polarimetry of a small number of brighter sources belonging to classes that were no more the cutting edge of X-Ray astronomy. Not surprisingly the originally foreseen polarimeters were excluded from both Einstein and Chandra.

Polarimetry was committed to hypothetic dedicated missions. Some were proposed but none advanced in the way to approval. 

\section{The Stellar X-Ray Polarimeter}
\label{sec:3}
The Stellar X-Ray Polarimeter (SXRP) was an important exception. The Spectrum Roentgen Gamma (SRG) mission was planned by the Space Agency and Academy of Sciences of Soviet Union, under the scientific leadership of Rashid Sunyaev, one of the scientists who contributed to the highest level to predict the relevance of X-Ray Polarimetry. SRG included many instruments in the band of soft and medium hard energy X-rays. The largest instrument was the Soviet Danish Roentgen Telescope (SODART), a pair of large area, medium optical quality, optics, with a focal length of 10m and with an effective area of the order of 1200 cm$^{2}$ each. Each telescope had a sliding platform in the focal plane capable to position different instruments. SODART by itself and, a fortiori, the whole satellite were so complex and massive that could host a polarimeter of around 50 kg mass, included rotation, without a major perturbation.
SXRP was based on an optimal use of both Bragg and Thomson technique. The difference in band allowed to stack the two stages. Around 10 centimeters  above the focal plane a flat graphite crystal at $45^{o}$ from the optical axis was diffracting 2.6 and 5.2 keV photons to a secondary focal plane. Since the diffraction follow the same laws of optical reflection the whole was still preserving the imaging properties of the telescope in a kind of a Newtonian mounting\cite{Silver1989}. The crystal was highly transparent to photons of higher energies that would impinge on a stick of lithium within a thin beryllium case. The two analyzers, namely the diffractor and the scatterer, were hosted in the center of a well made with four detectors\cite{Soffitta1998} . The detectors were proportional counters filled with a mixture of Xe, A and $CO_{2}$. The window was 150$ \mu $m thick except for a circular sub-window in the positon of the secondary focus for the diffracted photons, only 50$ \mu $m thick to leave a reasonable transparency for 2.6 keV photons. The detectors were position sensitive by exploiting the signal induced on a cathode plane subdivided according to a wedge and strip code. The background was minimized by an anticoincidence plane and by pulse shape discrimination. The whole instrument including the analyzers and the detectors was rotated around the optical axis. The rotation would perform the measurement with the modulation of the diffracted image in the auxiliary detector and to compensate systematics in the scattering stage. The Principal Investigator was Novic, replaced in a late stage in a late stage by Philip Kaaret. An Italian team was part of the project and contributed the detectors.

This was likely the best way to use the two conventional techniques in combination with a grazing incidence telescope. The telescope had a very large area and a band pass extended to 15 keV, namely with a reasonable overlap with the scattering cross section, so it was the best possible instrument with conventional (one-layer) optics. The main difficulty was the large tolerances in the alignment of the instrument with the focus of the telescope. This drove some design decisions with negative consequences. 

A diameter of the scatterer larger than required from the convergence of the X-ray beam and from the tolerance on positioning increased the self-absorption in the scatterer with a loss of signal and, most relevant, an increase of the systematics in case of offset from the axis. 

Also the distance of the detectors from the axis was larger than needed, to minimize the impact of possible offset of the lithium stick in the focus. But in a scattering polarimeter the background is proportional to the surface of the detectors. The predicted background rate would range from 200 to 400 mCrab\cite{Kaaret1994}. This means that the benefit of having an instrument in the focus of a telescope would only apply for the observation of a limited subset of brighter sources. The Bragg stage, thanks to the technique that preserves the imaging (although with a poor quality of the optics) would have a low background, virtually negligible. But taking into account all the losses of signal the effective area at 2.6 keV was of the order of 10 $cm^{2}$ on a bandwidth of the order of 100 eV. Polarimetry is a discipline in starvation of photons. A reasonable MDP could be achieved only for sources of tens of mCrab. In the baseline program for SRG the share of observing time with SXRP as a prime was of 5$ \% $. So in practice a few brighter sources would be probed in the first years and the extragalactic sky was out of reach unless in the case of an extreme flare. In any case SXRP would perform polarimetry of a certain number of bright galactic sources down to the level of a few $\%$, and this would be a serious step forward with respect to OSO-8. 

The experiment was completed, integrated and fully calibrated. It passed all the acceptance tests and was packed waiting to the delivery to be integrated in the SRG payload. Unfortunately the making of SRG was slowed and eventually stopped by the general sinking of the Soviet System. The fact that the end of the program was never declared had the effect that SXRP in a perpetual floating situation acted as a stopper for any other proposed X-ray polarimeter, as, for instance, for the one proposed for XMM.

One interesting heritage of SXRP was the calibration. It was performed at Lawrence Radiation Laboratory in Berkley, by means of X-ray generators and crystal diffractors. A particular attention was devoted to the calibration of the systematics foreseen for SXRP, such as the off-axis or a slight inclination. A special attention was spent, with a large investment of time, to disentangle the intrinsic polarization of the calibrator from the measurement of the small effects of spurious modulation intrinsic to the instrument. From the point of view of data analysis the technique of Fourier Analysis was widely used, with very promising results taking into account a starting point with a high level of systematics\cite{Tomsick1997}.

\section{The quest for photoelectric polarimeter}
\label{sec:4}

The photoelectric effect had been studied in laboratory in the $'$20s essentially with two methods. The first consisted in sending X-rays on a thin solid target and collect the emitted electrons with a detector on a goniometer geometry.  The method has a fundamental difficulty for astrophysical applications. The modulation strongly depends on the penetration of the photon and this depends on the energy. In a laboratory set up the measurement can be performed with photons of known energy. In an astrophysical measurement the energy is unknown and the result is of difficult interpretation. 

The other approach is to have a detector where the whole energy is lost in active regions. In the laboratory this was performed with excellent imaging capability with cloud chambers. The photographic read out was providing a very good image of the tracks, with all the details. The quest for a photoelectric polarimeter for space is the search for a device with quality as similar as possible to a cloud chamber but self-triggered and suited to be hosted in a focal plane.
 
It is worth to mention the fact that for a short period the SXRP Team explored the possibility to introduce a new device based on  the first of the two concepts described above. Notwithstanding a relatively poor modulation factor of the order of 25$\%$, the full exploitation of the optics at low energies and the condition to be background-less down to very weak sources, promised an estimated sensitivity that would enable polarimetry of extragalactic sources. Further measurements showed that the actual modulation was much lower\cite{Hanany1993} and this photoelectric polarimeter was no more adopted. But this showed to the small community interested on this topic how relevant would be the implementation of a photoelectric imaging device for the focal plane. 

SXRP was a big hope for X-ray polarimetry but, as the mission was continuously delayed and never stopped, it became a stopper for any new proposal. Every proposal, such as that submitted for XMM, would face the statement that we should wait for the results of SXRP first.
In this phase of suspense, the idea that the next step would be a dedicated mission based on photoelectric effect started to spread. In the very beginning of X-Ray Astronomy some laboratories tried to identify a method to perform polarimetry based on the photoelectric effect. Given the limited imaging capability of typical wire chambers, these were based on the search of modulation of some macroscopic parameter deriving from the asymmetric angular distribution of the photoemission. Riegler at GSFC\cite{Riegler1970} searched for a modulation with phase of coincident signals in nearby anodes for photons entering parallel to the wires. Sanford at MSSL\cite{Sanford1970} searched for a difference of the rise-time of tracks parallel or perpendicular to the anode, in a proportional counter. Both results were not conclusive and no further efforts were done after Novick results indicated the Bragg as the most effective technique.

With ASCA, Chandra and XMM-Newton the CCD took the place of Gas Proportional Counter as tow horse of X-Ray Astronomy. Measurements showed a large majority of single hits but also a certain number of double hits, especially at higher energies. The fluorescence yield for Si is only 3$\%$. These hits were interpreted as a transfer of charge to the nearby pixel from photoelectrons created at a distance from the edge lower than the range. Given the angular distribution of the photoelectrons the double hits along rows and those along columns have memory of the polarization of the X-rays. This was verified by measurements and simulations by Osaka University\cite{Tsunemi1992}, Max Planck\cite{Schmidt1996} and Leicester\cite{Holland1995} teams. The method was obviously attractive because it does not require a dedicated instrument. The measurement of polarization is the byproduct of the measurement of position and energy. In truth this is only apparently true. The range of the electrons is much shorter than the pixel size, so that the efficiency is very poor except at higher energies (>10 keV) where the efficiency of the optics is typically low and the chip is transparent to X-Rays. Moreover the square pattern of the sensor is a problem itself. A beam polarized at $45^{o}$ from the array and a beam unpolarized give exactly the same pattern and the system to work would need a rotation of the satellite or more detectors angularly misaligned. But the most serious difficulty is that in these conditions the probability to have a double hit depends more on the distance of the absorption point from the edge then on polarization and with a point spread function that varies very sharply on the detector pattern the result would change completely for aspect change of seconds of arc.

For basic reasons the implementation of the photoelectric method was only possible with gas. The team of Marshall Space Science Laboratory lead by Brian Ramsay made an important step forward with a gas detector with an absorption region with an electric field to drift the electrons of the track to a grid acting as a multiplication plane. The gas mixture was based on Argon but included the TAE, a luminescent component. During the multiplication optical photons were emitted so that the track became somehow luminous. An optics would focus an image of the track on a CCD. 70 years after the images from cloud chambers these were the first images of photoelectron tracks.  They were acquired by optical methods, just like the cloud chamber experiments, with a CCD, that was taking the place of photographic films. The method was very promising but the minimum energy was of 20 keV\cite{Austin93}. At the time the multi-layer optics were in a pioneering phase so that the path toward a realistic experiment was still to be defined.

But the technology of gas detectors was passing a season of great improvements. The impetuous development of microelectronics also made feasible detection patterns with fine subdivision and all the simulations showed that a 2-dimension pixel imager would be the full exploitation of the photoelectric effect.

\section{The first gas pixel detectors}
\label{sec:5}

The team of IAPS (institutionally evolved from CNR to INAF), lead by the author and by Paolo Soffitta, another component of the SXRP collaboration, started a collaboration with the team of INFN-Pisa lead by Ronaldo Bellazzini, an outstanding personality in the field of gas detectors for particle physcs. This INAF/INFN partnership could benefit of the convergence of two traditions and eventually succeeded to arrive to the first detector suited for the purpose. 

A first testing on one dimension was performed with a microgap detector with a pitch of 200$ \mu $m filled with a Ne-DME mixture at 1 atmosphere pressure. Data showed that the track length of electrons from photons of 5.4 polarized perpendicular to the strips were definitely longer than tracks from photons parallel\cite{Soffitta1995}. To my knowledge this is the first unambiguous detection of an effect in the classical range 2-8 keV, viable for polarimetry. The key was the use of a gas of low atomic number and the high spatial resolution, although in one dimension only. Moreover the data were consistent with simulations software performed starting with a program for micro-analysis, given that at the epoch, general purpose software of the GEANT family were not yet adequate below 10 keV.  

Encouraged from this result the INFN/INAF team made a first prototype that was the real turning point of photoelectric Polarimetry from a wishful thinking to a practical implementation. 
 
The development of the GPD was articulated into various steps with some substantial commonality:

A gas cell with a conductive window.

An electric field parallel to the optical axis to drift the electrons of the track to a multiplication stage.

A Gas Electron Multiplier (GEM) to perform a proportional multiplication of the electrons to amplify the track while preserving the shape. In a first generation of devices the signal from the GEM gave the trigger for the data acquisition.

A sense plane of metal pads distributed on a honeycomb pattern to act as anodes to collect the charge multiplied in the gas above the pad itself.

A dedicated front end electronic chain independent for each pad.

An important driver of this design was the search for maximum axial symmetry, aimed to prevent any possible systematics and hopefully to avoid the need for rotation that till then was a mantra of X-ray polarimetry and one of the source of complications that made of polarimetry an odd companion in focal planes.

The first prototype based on trigger by GEM and collection plane built with multi-layer printed circuit technology. The pad pitch was 150$ \mu $m. The GEM pitch 60$ \mu $m. Signals of each pad were routed horizontally to the input of an ASIC chips, relatively far from the gas cell. 
The detector was filled with Ne and DME as a quencher. The signal from the GEM was used to trigger the ASIC data acquisition. Eventually the charge collected from each pixel was fetched to the output and A/D converted. Photons of 5.4 and 5.9 keV generated ionization tracks in the gas. From the analysis of the track images the direction of the emission was derived. The histogram of angles was flat for unpolarized 5.9 keV photons and modulated for 5.4 keV photons, which had been polarized by scattering at 90$ ^{o} $ on a lithium target\cite{Costa2001}. Although the statistics was not very high the papers presenting these results was the re-start of activities on X-ray polarimetry that can be classified into three groups.
1)	An activity to improve and optimize the performances of prototypes and to build a real detector that could be the core of a space experiment
2)	A rejuvenation of theoretical analysis to predict the potentiality of this new observable
3)	The proposition of missions of polarimetry at every announcement of opportunity or in the context of large multi-instrument missions.
The results from the prototype demonstrated that the basic physics was correct but, for a realistic proposal, some problems had to be solved. The major limits of the prototype were:
1)	The multi-layer technique could allow for a limited number of pixels, of the order of one thousand.
2)	 The pads could not be smaller than 100$ \mu $m
3)	The routing of the charge to chips far from the gas cell on an horizontal lay-out were the source of a noise much larger than that intrinsic to the anode read-out.
The total encumber of the detector and electronics, although already much better than that of the conventional experiments, was still relatively large compared with the small active sensing surface. 
Notwithstanding these limitations at the Announcement of Opportunity from NASA for a Medium Size Explorer a Mission named INSPIRE was proposed including two telescopes one with a a Microcalorimeter.

The second step was the inclusion of the whole electronics in an ASIC chip. The upper layer of the chip was the array of pads in hexagonal pattern. The analog electronics was in the lower layers under the projection of the pad. The signals were hold on the trigger from the GEM and routed to the output. Two generations of these ASIC were made in sequence. The first one had 2100 pixels, the second one 22000\cite{Bellazzini2004}, in an hexagonal pattern. The pixel size arrived to 80$ \mu $m.The most impressive evolution was to collapse the whole ingredients of the detecting system into a small volume around the active cell of gas. This was basically different from other micro-pattern prototypes ad was named the Gas Pixel Detector (GPD).

The following evolution was a further decrease of the pixel size down to 50$ \mu $m and the increase of the number of pixels to 104000. With the previous philosophy to trigger on the GEM and A/D convert the charge of each pixel dead time would have diverged. The last step was the addition of self-triggering capability\cite{Bellazzini2006}.  The ASIC was therefore playing different roles: the bottom of the detector, the sensing multi-pad plane, the front end electronics, the triggering electronics. The chip was mounted on the bottom of a sealed ceramic case with a beryllium window and the whole weighted 80g. In practice it was very close to what is needed for a space mission. 

The chip could be used in principle with different gas mixtures, pressures and different thicknesses of the drift/absorption gap. To operate in the 2-8 keV band Ne with Dymethylether (DME) quenching was used. Following computation and tests a filling with DME and a 20$ \% $ of He was found slightly more performing. Eventually pure DME was used, although some problems could be expected with this chemically aggressive substance. As a quencher and even more as the main detecting gas DME is outstanding from the point of view of diffusion. Beside the choice of the gas mixture the pressure and the thickness of the absorbing cell was object of optimization. A pressure around one atmosphere allowed for easier operation both in air and in vacuum and could be contained with a 50$ \mu $m thick Beryllium window. The ingredients for the trade-off are:

\begin{enumerate}
\item{The thickness determines the efficiency.}
\item{The thickness determines the blurring of the track due to lateral diffusin in the drift.}
\item{The thickness determines a  blurring of the image in the focal plane of the optics.}
\end{enumerate}

Point 1) is relevant because Polarimetry is a photons starving technique. 
Point 2) impacts on the modulation factor, especially at lower energies. 
Point 3) is relevant only for optics of good quality of the order of 10 arcseconds or better. In fact the inclination angles are fixed by the band of the instrument and for medium quality telescopes (from 0.5 to  2 arcminutes) the defocusing effect is significantly minor in comparison with the telescope p.s.f..

The optimization of course depends on the band of optics and on the spectrum of a particular source. The optimization for a typical optics of 4 m focal length and a spectrum E$ ^{-2} $ was a mixture of 80$\%$ DME and 20$\%$ He at one atmosphere. Various  sealed GPD with this filling and a window of 50 micron of Beryllium were built, studied, and tested to evaluate the performances included the resistance in a space environment. This instrument was already mature for an actual mission and was proposed for various announcements to various Agencies (NASA, ESA, ASI, CAS).

\section{The Time Projection Chamber }
\label{sec:6}
At the NASA Announcement of Opportunity for a SMEX in 2003 a Team of Goddard Space Flight center lead by Jean Swank and based on the Laboratory for High Energy Astrophysics lead by Keith Jahoda and the Italian Teams that had developed the GPD lead by Ronaldo Bellazzini and Enrico Costa submitted a proposal for an Advanced X-Ray Polarimeter. AXP was based on three conical optics like those for ASTRO-E2 three GPDs in the focus. The proposed mission was not selected but NASA granted the LHEA with funds for the development of their own detectors for photoelectric polarimetry. This was the Time Projecton Chamber (TPC)\cite{Black2007}. 
The process to optimize the performance of the GPD had shown the limitation in the axis-symmetric design. The trade-off between efficiency and modulation factor even in mixtures with minimum diffusion was achieved at values around 10-15$\%$ of efficiency at a peak of 2-3 keV. On the other side in the GPD the ASIC must be orthogonal to the axis. This forbids to recover the efficiency by stacking more GPDs.

The physics base of LHEA detector was the same of GPD and of any other modern photoelectric polarimeter, namely to image the ionization track produced by a photoelectron in a gas detector. But in order to achieve a higher efficiency they decided to drop the axial symmetry. In the TPC the electrons of the track are drifted on one side so that the drift path and the associated blurring is kept at an acceptable level. The thickness of the absorption gap can be made very long in order to to increase the efficiency, potentially to values close to 1.  At the end of the drift a GEM, just like in the GPD, amplifies the track. The multiplied electrons are collected by a set of vertical strips parallel to the optical axis. The other dimension, namely that of the drift field, the track is measured by the Time Projection method. Given that the the two coordinates of the images are derived with different methods and the drift method scales with the knowledge of the drift velocity an additional care is needed. The track is imaged but, as usual in drift detectors, when a photon is absorbed no signal gives the start of the drift time. The track is imaged, and the angular direction of the photoelectron can be measured,  but the absorption point cannot be derived. This impacts negatively in two ways:
\begin{itemize}
\item{Image resolved polarimetry cannot be performed}
\item{The reduction of background for the use of the optics is limited to one dimension only. .}
\end{itemize}
These limitations are intrinsic to the method.
Some other problems were associated to the specific implementation. These include a large dimension that, combined with the need of rotation, made it more cumbersome and a relatively limited band width. Conversely the TPC could be proposed from early times as a medium energy polarimeter. In fact the lateral read-out was compatible with both a thick absorption gap or a thin absorption gap with a back window that would allow the higher energy photons to reach another polarimeter mounted in series (e.g. a scatterig polarimeter).

To summarize the TPC is more efficient of the GPD but is not imaging. The GPD should have a lower background. Moreover the TPC needs rotation to compensate for systematics while the GPD, with hexagonal pads, by construction, at the first order could be free from systematic effects. Actually in the implementation for IXPE sysiematics of the order of 1$\%$ at lower energies were deteced and calibrated.
The constructive details and the performances of the GPD and the TPC are discussed in another part of this Handbook. Here I give some hint on the following history.

\section{Toward a mission }
\label{sec:6}

With the implementation of GPD and the alternative development of TPC X-Ray Polarimetry had reached the maturity to compete in the selections by the space agencies. In theory, beside this baseline range of 2 - 8 keV,  other  ranges could be covered, such as the soft band or the hard X-Rays. But  in this classic band a large set of results could be expected for most classes of X-ray sources, on the basis of theoretical analysis and achievable sensitivity. No comparable wealth of results was to be expected for other band, that would also require more challenging instrumental efforts. This limited the path toward a mission to a comfrontation of GPD and TPC both filled with low Z gas.

This band was well matched with that of XEUS, the Large X-Ray mission studed for years by ESA. A GPD polarimeter was added and was part of the baseline focal plane payload in all the various configurations of XEUS and IXO\cite{Costa2008}. 

But the first implementation of this new technique could not be committed to such a remote horizon.
An Announcement Of Opportunity for Two Small Scientific Satellites was issued by the Italian Space Agency (ASI) on 2007. A team lead by Enrico Costa, Ronaldo Bellazzini and Gianpiero Tagliaferri proposed POLARIX a mission with three telescopes, residual of another experiment foreseen for the SRG mission and three GPD\cite{Costa2010}. The mission was selected for a phase A study and was ranked second in the following selection. But eventually all the program of Small Scientific Missions was discontinued by ASI. 

After the development of  TPC the AXP collaboration splinted toward competing proposals. At the further SMEX Announcement of Opportunity on 2007 the GSFC team, lead by Jean Swank proposed GEMS\cite{Swank2010}, based on the TPC, while the MSFC team lead by Martin Weisskopf with the Italian Team proposed the Imaging X-Ray Polarimetry Explorer\cite{Weisskopf2008b}  based on the GPD. Both missions were based on three telescopes and a deployable boom. NASA selected GEMS for an advanced study and eventually for flight. But in 2012 GEMS (that meanwhile had been descooped to two telescopes) was discontinued by NASA for programmatic reasons. For the second time a mission approved and eventually not launched acted as a stopper to other proposals.

After the suppression of GEMS instruments aimed to X-ray polarimetry were proposed again. A descoped version of POLARIX was proposed at the first ESA Anouncement for a small mission\cite{Soffitta2013b}.
The most advanced proposal was the X-ray Imaging Polarimetry Explorer proposed by an European team lead by Paolo Soffitta to the M4 AOO of ESA\cite{Soffitta2013b}. XIPE was selected for a phase A study together with two other missions. XIPE was based on three telescopes with GPDs in the focus. XIPE was similar to IXPE with a collecting area of the 3 mirrors around twice that of IXPE. 
 
On the other side of the Ocean at the following AOO for a SMEX in 2014 many projects of X-Ray polarimetry were submitted and two of them were selected for an A phase study. IXPE was a rejuvenated version of the homonym precursor with a major role of the Italian collaboration that would provide the whole instrument\cite{Soffitta2021}  in the focal plane of three telescopes built by the MSFC team. Martin Weisskopf was the PI and Paolo Soffitta the Italian PI. Praxys was based on the GEMS design, with two telescopes (following the GEMS design) and TPCs in the focus. Keth Jahoda was the PI of this other proposal.
At the end of the phase A study in 2017 NASA selected IXPE for flight.  As a consequence XIPE was no more considered for the ESA M4 selection.

In parallel to the development of the hardware the analysis tools and the statistical context were better defined by Martin Weisskopf\cite{Weisskopf2010} and Stroheymer and Kalmann\cite{Strohmayer2013}. The use of Stokes Parameters for the analysis was better defined. Interestingly it was proposed to apply the Stokes formalism also to single photons\cite{Kislat2015}. With management by NASA and ASI, IXPE passed all the program stages and was launched on December 9 2021\cite{Weisskopf2021}.

\section{Not only IXPE}
\label{sec:7}
IXPE, launched on 2021 is performing as hoped and planned. During the years needed to have this mainstream experiment in orbit a certain level of activity was performed. 

The activity of polarimetry in the hard X-Rays was performed with stratospheric balloons and was somehow decoupled from the activity at lower energies onboard satellites. The two main experiments have been POGO and X-Calibur. Both are scattering polarimeters but conceptually quite different. POGO is a collimated detector. The scattering element is plastic scintillator. The background is reduced with an heavy passive shielding and with an active anti-coincidence including active collimator. POGO evolved from a first design (POGOLite)  to a larger version POGO+\cite{Chauvin2016}. Scientific results include an observation of the Crab\cite{Chauvin2017} and of CygX-1\cite{Chauvin2019}. X-CALIBUR is based on an optics with 12m focal length and, in the focus, a scattering stick of Beryllium surrounded with an absorption well made of four strips of CZT detectors. X-Calibur was launched from Antarctica. It observed GX301-2 in a flare status and found an upper limit to polarization\cite{Abarr2020}. Following the results, the project evolved to an advanced version named XL-Calibur, with special attention is paid to improve the background shielding\cite{Abarr2021}. In general scattering polarimeters have larger background partially compensated with a higher  modulation factor on a wider energy band. This will be greatly improved when these instrumets will be hosted on satellites, allowing for a broader band and suitable observing time. 

Also in the domain of Hard X-Rays and at the edge of the range of interest of this chapter a small satellite devoted to polarimetry is expected to be launched from ISRO in the next years. POLIX is constituted of collimators, a Beryllium scatterer and proportional counters\cite{Paul2016}l, heritage of the ASTROSAT design.

Three years before the launch of IXPE Polarlight, a cubesat with a GPD based on the same ASIC of IXPE and a collimator was launched on a sun synchronous orbit by Hua Feng of Tsing Hua University\cite{Feng2019}. Polarlight by long integrations measured again after 40 years the polarization of the Crab\cite{Feng2020}, with a possible change of angle after a glitch,  and of Sco X-1\cite{Feng2022}.   Interestlngly PolarLight measured a background rate at a level of one 20th of Crab. This shows that a large number of galactic sources could be observed with collimated photoelectric detectors without the programmatic and financial loads of optics.  On the opposite approach the enhanced X-ray Timing and Polarimetry mission(eXTP), in an advanced state of approval, includes 4 telescopes devoted to polarimetry\cite{Zhang2019}. The general feature are similar to those of IXPE but the collecting area is 4 timeslarger and the observations will benefit of the simultaneous measurement with high throughput spectroscopy and timig instruments.

While the path toward new data was so slow and painful, some progress was done in the development of methods of analysis. These include a better understanding of the parameters defining the sensitivity, more clear use of the Stokes Parameters, and the development of Bayesian algorithms.  POGO and PolarLight used for the first time the Bayesian analysis. 

The methods of analysis are the subject of next chapters.

\bibliographystyle{elsarticle-num}
\bibliography{References-3}
\end{document}